\documentclass[twocolumn,floats,amsmath,amssymb,prb]{revtex4}

\usepackage{graphicx}

\begin{document}

\title{Configurational entropy of ice from thermodynamic integration}
\author{Carlos P. Herrero}
\author{Rafael Ram\'{\i}rez}
\affiliation{Instituto de Ciencia de Materiales de Madrid,
         Consejo Superior de Investigaciones Cient\'{\i}ficas (CSIC),
         Campus de Cantoblanco, 28049 Madrid, Spain }
\date{\today}

\begin{abstract}
The configurational entropy of ice
is calculated by thermodynamic integration from high to low
temperatures. We use Monte Carlo simulations with a simple
energy model which reproduces the Bernal-Fowler ice rules.
This procedure is found to be precise enough to give
reliable values for the residual entropy $s_{th}$ of different ice 
phases in the thermodynamic limit.
First, we check it for a two-dimensional ice model.
Second, we calculate $s_{th}$ for ice Ih,
and compare our result with those previously given in the literature.
Third, we obtain $s_{th}$ for ice VI, for which we find a value 
clearly higher than for ice Ih. 
\end{abstract}

\maketitle

\section{Introduction}

Water is known to show a wide variety of solid phases, and
in fact, sixteen different crystalline ice phases have been 
identified so far.\cite{du10,ba12}
The determination of their crystal structures and
stability range in the pressure-temperature phase diagram has been
a matter of research for several decades.
However, despite the large amount of experimental and theoretical work on
the solid phases of water, some of their properties still lack a complete
understanding. This is mainly due to their peculiar structure, where 
hydrogen bonds between contiguous molecules give rise to
properties somewhat different than those of most known liquids and
solids.\cite{ei69,pe99,ro96}

In all ice phases (with the exception of ice X),
water molecules appear as well defined entities forming
a network connected by H-bonds. In this network each
water molecule is surrounded by four others in a more or less distorted
tetrahedral coordination. The orientation of each molecule with respect
to its four nearest neighbors fulfills the so-called Bernal-Fowler ice 
rules. These rules state that each H$_{2}$O molecule is oriented 
in such a way that its two protons point toward adjacent oxygen atoms 
and that there must be exactly one proton between two contiguous oxygen
atoms.\cite{be33} In the following we will refer to these rules simply
as ``ice rules.''

The presence of orientational disorder in the water molecules is
a property of several ice phases. While the oxygen atoms show a
full occupancy ($f$) of their crystallographic positions, the
hydrogen atoms may display a disordered spatial distribution as 
indicated by a fractional occupancy of their lattice sites. 
Thus, ice Ih, the stable phase of solid water under normal conditions,
presents full proton disorder compatible with the ice rules, 
i.e., occupancies of H-sites of $f = 0.5$. 
However, other phases such as ice II are H-ordered, while others as 
ice III are characterized by a partial proton ordering, 
i.e., some fractional occupancies of H-sites are different from 0.5. 

Configurational disorder of protons in ice was first studied
by Pauling,\cite{pa35} who estimated its contribution to the
entropy of a crystal of $N$ molecules to be
$S = N k_B \ln (3/2)$.
This combinatorial estimate turned out to be in good agreement with 
the ``residual'' entropy derived from the experimental heat capacity 
of ice Ih,\cite{gi36,ha74} although the calculation did not take into 
account the actual structure of ice Ih.
Nagle\cite{na66} calculated later the residual entropy of hexagonal 
ice Ih and cubic ice Ic by a series method, and found in both cases very 
similar values, which turned out to be close to but slightly higher than 
the Pauling's estimate. 
More recently, Berg {\em et al.}\cite{be07} have used multicanonical 
simulations to calculate the configurational entropy of ice Ih, assuming 
a disordered proton distribution compatible with the ice rules.
Moreover, several authors have calculated the configurational entropy of 
partially ordered ice phases,\cite{ho87,ma04b,be07b}  
a question that will not be addressed here.

Ice-type models are important not only for condensed phases
of water, but also for other kinds of materials showing atomic
disorder,\cite{is04} as well as in the statistical mechanics
of lattice models.\cite{zi79,be89}
Although exact analytical solutions for the ice model in the actual
three-dimensional ice structures are not known at present,
an exact solution was found by Lieb for the two-dimensional 
square lattice.\cite{li67,li67b} In this case, the
configurational entropy results to be $S = N k_B \ln W$, with  
$W = (4/3)^{3/2} = 1.5396$, somewhat higher than the Pauling
value $W_P = 1.5$.

In this paper, we present a simple, but ``formally exact'' method, to
obtain the configurational entropy of H-disordered ice structures.
It is based on a thermodynamic integration from high to low
temperatures, for an ice model which reproduces the ice rules
at low temperatures.

\section{Computational method}

To calculate the configurational entropy of the different ice
structures, we consider a simple model. The only requirement for this
model is that it has to reproduce the ice rules at low temperature. 
Thus, irrespective of its simplicity, it can give the
actual entropy if an adequate thermodynamic integration is carried out. 

For concreteness, we summarize the ice rules:\cite{be33}

(1) There is one hydrogen atom between each pair of neighboring 
oxygen atoms, forming a hydrogen bond.

(2) Each oxygen atom is surrounded by four H atoms, two of them 
 covalently bonded and two other H-bonded to it.  

Our model is defined as follows. We consider an ice structure as defined 
by the positions of the oxygen atoms, so that each O atom has four
nearest O atoms. This defines a network, where the nodes are the O
sites, and the links are H-bonds between nearest neighbors. 
The network coordination is four, which gives a total of $2N$ links,
$N$ being the number of nodes. We assume that on each link there is
one (and only one) H atom, which can move between two positions 
on the link (close to one oxygen or close to the other).

\begin{figure}
\vspace{0.3cm}
\hspace{-0.2cm}
\includegraphics[width=8cm]{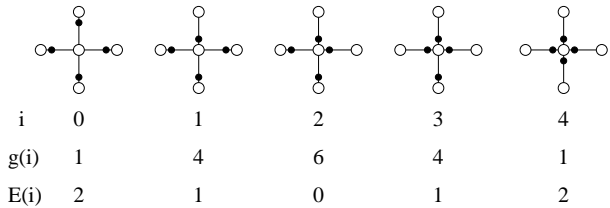}
\caption{
Sketch of the different hydrogen configurations around an
oxygen atom. Open and solid circles represent oxygen and
hydrogen atoms, respectively.
For each configuration, $i$ is the number of H atoms close
to the central oxygen, $g(i)$ indicates its multiplicity,
and $E(i)$ refers to its energy in the model described in the
text.
}
\label{f1}
\end{figure}

Given a configuration of H atoms on an ice network, the energy $U$ is 
defined as:
\begin{equation}
U = \sum_{n=1}^N  E(i_n)
\end{equation}
where the sum runs over the $N$ nodes in the network,
and $i_n$ is the number of hydrogen atoms covalently bonded to the
oxygen on site $n$,
which can take the values 0, 1, 2, 3, or 4.
The energy associated to site $n$ is then $E(i_n) = |i_n - 2|$ 
(see Fig.~1), having a minimum for $i_n = 2$, which 
imposes the fulfillment of the ice rules at low temperature.
In this way, all hydrogen configurations compatible with the ice
rules on a given structure are equally probable in the low-temperature 
limit, i.e., it is implicit in the model that all configurations obeying 
the ice rules have the same energy.

We note that this simple model, although it is a convenient tool for 
our present purposes, does not represent any realistic interatomic
interaction, in the sense that we are not dealing with a real ordering 
process, but with a numerical approach to ``count'' H-disordered
configurations of ice. Since the entropy is a state function, one can 
obtain the number of configurations compatible with the ice rules by a 
kind of thermodynamic integration from a reference state ($T \to \infty$)
for which the H configuration is random (does not respect the
ice rules), to a state in which these rules are strictly
fulfilled ($T \to 0$). 
We note that in our calculations we use reduced variables,
so that all quantities such as the energy $U$ and the temperature 
$T$ are dimensionless. The entropy per site $s$ that we calculate
is therefore related with the physical configurational entropy $S$
as $S = N k_B s$.

The heat capacity per site, 
\begin{equation}
  c_v(T) = \frac{1}{N} \frac{d \langle U \rangle}{d T}
\label{cvt}
\end{equation}
has been obtained from the energy fluctuations at temperature $T$, 
by using the expression\cite{ch87}
\begin{equation}
c_v(T) = \frac {(\Delta U)^2} {N T^2}  \,  ,
\end{equation}
where $(\Delta U)^2 = \langle U^2 \rangle - \langle U \rangle^2$.
The configurational entropy per site can be obtained from the heat
capacity by thermodynamic integration as
\begin{equation}
 s(T) = s(\infty) + \int_{\infty}^T  \frac{c_v(T')}{T'} \, dT'
\label{entrop1}
\end{equation}
In our case, the entropy per site for $T \to \infty$ is given by
\begin{equation}
  s(\infty) = \frac{1}{N} \ln (2^{2 N}) = 2 \ln 2 
\label{entinfi}
\end{equation}
where $2 N$ in the exponent indicates the number of links in the 
network under consideration (Note that $2^{2 N}$ is the total number
of possible configurations in our model, since each of the $2 N$
links admits two different positions for an H atom).

A practical problem with Eq.~(\ref{entrop1}) in a thermodynamic
integration is that the limit $T \to \infty$ cannot be reached, and
any cutoff in the temperature, even if this is taken at large $T$, 
can introduce systematic errors in the calculated entropies. 
To overcome this problem we use the fact that an analytical model 
can approximate very well the high-temperature thermodynamic
variables, with an error smaller than the error bars associated to
the simulation procedure.
Such an analytical model can be obtained by considering the nodes 
as ``independent'', as in Pauling's original calculation for a
hypothetical network including no loops.
In this case, the partition function is given by 
\begin{equation}
  Z =  \frac {z^N} {2^{2N}}
\label{zz}
\end{equation}
where $z$ is the one-site partition function: 
\begin{equation}
  z =  6 + 8 \, {\rm e}^{-1/T} + 2 \, {\rm e}^{-2/T}  \, ,
\label{z68}
\end{equation}
(as derived from the Boltzmann factors for 16 possible configurations of
four hydrogen atoms; see Fig.~1)
and the term $2^{2N}$ in the denominator appears to avoid counting links
with zero or two H atoms. In fact, the high-temperature limit of $Z$
in Eq.~(\ref{zz}) is $Z_{\infty} = 16^N / 4^N = 2^{2N}$, which
is the number of configurations compatible with the condition of
having an H atom per link.
The energy per site is
\begin{equation}
 \langle u \rangle = \frac{4}{z} \left( 2 \, {\rm e}^{-1/T} + {\rm e}^{-2/T} 
              \right) \, .
\label{eindep}
\end{equation}

For the actual ice structures, the corresponding oxygen networks 
contain loops, which means that the factorization in the partition
function in Eq.~(\ref{zz}) is not possible.  However,
at high temperatures thermodynamic variables for the real structures
converge to those of the independent-site model. 
At high temperature, $\langle u \rangle$ in Eq.~(\ref{eindep}) 
can be expanded in powers of $1/T$ so that
\begin{equation}
\langle u \rangle = \frac34 - \frac{7}{16 \, T} + \frac{3}{64 \, T^2} +
      \frac{19}{768 \, T^3} + ... 
\end{equation}
and using Eqs.~(\ref{cvt}) and (\ref{entrop1}) we find for the entropy
\begin{equation}
  s(T) = s(\infty) - \frac{7}{32 \, T^2} + \frac{1}{32 \, T^3} +
       \frac{19}{1024 \, T^4} + ...
\end{equation}
Keeping terms for $s(T)$ up to $1/T^4$, we find 
$s(T = 10)$ = 1.38414.
We have checked that at temperatures $T \sim 10$, both the energy and 
heat capacity derived from this model coincide (within error bars)
with those derived from our Monte Carlo simulations for the actual
networks studied here.
Thus, to obtain the configurational entropy for the ice structures, 
our thermodynamic integration in fact begins at $T = 10$, a
temperature at which the entropy of the actual network is taken as
that of the analytical (no loops) model.
We note that at lower temperatures ($T \sim 1$), the simulations for the 
considered networks yield energy values different from those derived
from the ``independent-node'' model in Eqs.~(\ref{zz}) and (\ref{z68}), 
which is in fact the reason why we find for these networks entropy values 
different from the Pauling result.

With this simple scheme for the energy, we have carried out
Monte Carlo simulations on ice networks of different sizes.
The largest networks employed here included 
3600, 2880, and 3430 sites for the square lattice, ice Ih, and
ice VI, respectively.
Periodic boundary conditions were assumed.
Sampling of the configuration space was carried out by the Metropolis
update algorithm.\cite{bi97}
For each network we considered 360 temperatures in the interval
between $T$ = 10 and $T$ = 1, and 200 temperatures in the range from
$T$ = 1 to $T$ = 0.01.
For each considered temperature, the simulation started from the 
last hydrogen configuration in the previous temperature, and then
we carried out $10^4$ Monte Carlo steps 
for system equilibration, followed by $8 \times 10^6$ steps for
averaging of thermodynamic variables. Each Monte Carlo step consisted 
of an update of $2 N$ (the number
of H-bonds) hydrogen positions successively and randomly selected.
Finite-size scaling was then employed to obtain the configurational
entropy per site $s_{th}$ corresponding to the thermodynamic limit
(extrapolation to infinite size).

Using a simple energy model as that employed here, an alternative
procedure to calculate the configurational entropy can consist in
obtaining directly the density of states as a function of the energy,
as described elsewhere for order/disorder problems in condensed 
matter.\cite{he92}
Then, in our case the entropy could be obtained from the number of states 
with zero energy, i.e., compatible with the ice rules.

\section{Results and discussion}

We have applied the method described above to calculate the
configurational entropy of the ice model in three different
networks. First, we check the precision of the method for the
two-dimensional square lattice, for which an exact analytical solution 
is known. Then, we calculate $s_{th}$ for the familiar ice Ih,
and compare our results with those obtained in earlier work.
Finally, we present results for the configurational entropy of ice VI, 
for which one expects $s_{th}$ to be appreciably different 
from ice Ih, due to the presence of four-membered rings of water 
molecules in its structure. 

\begin{figure}
\vspace{-1.1cm}
\hspace{-0.5cm}
\includegraphics[width= 9cm]{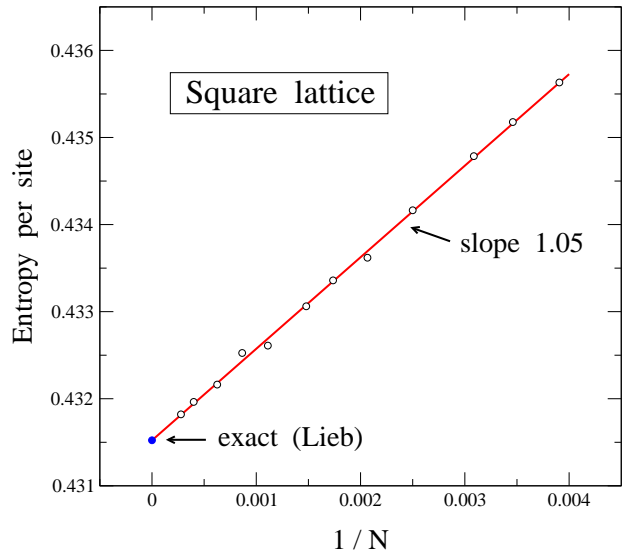}
\vspace{-0.7cm}
\caption{
Entropy per site as a function of the inverse lattice size ($1/N$)
for
the two-dimensional square lattice.
Open circles indicate results of our thermodynamic integration in the
limit $T \to 0$.  Error bars are in the order of the symbol size.
A solid circle shows the exact analytical result obtained by
Lieb.\cite{li67}
}
\label{f2}
\end{figure}

For each considered network, we have obtained the configurational
entropy $s_N$ in the limit $T \to 0$ for several system sizes $N$, 
as described in Sect.~II.
In Fig.~2 we present $s_N$ for the ice model on the two-dimensional 
square lattice.
Open symbols represent the configurational entropy derived 
from our thermodynamic integration, as a function of the inverse 
system size. 
We find that $s_N$ decreases for increasing system size $N$, and in
fact there is a linear dependence of $s_N$ on $1/N$ for 
$N \gtrsim 150$, in the form 
\begin{equation}
  s_N = s_{th} + \frac{a}{N} \, ,
\label{sns}
\end{equation}
where $a$ is a network-dependent parameter. 
For smaller system sizes (not shown in the figure), 
$s_N$ deviates slightly from the linear behavior, becoming smaller
than the value predicted from a linear fit for $N > 150$ sites.
Extrapolation of the linear fit for $1/N \to 0$ gives a value
$s_{th}$ = 0.43153(3), in good agreement with the exact solution 
for the square lattice found by Lieb\cite{li67} by the transfer-matrix
method: $s_{th} = \frac32 \ln (4/3) = 0.43152$.
Information on the least-square fit carried out here is given in Table~I. 

\begin{table*}
\caption{Entropy for the ice model on the square lattice, ice Ih, and
 ice VI, as derived from our Monte Carlo simulations.
 $n_P$: number of data points employed in the linear fits;
 $a$: slope of the linear fit as in Eq.~(\ref{sns});
 $\rho$: correlation coefficient.
 For comparison, we give the entropy values obtained in earlier
works. }
\label{tab:2}
\vspace{5mm}
\begin{tabular}{c c c c c c c}
 Ice  & Author &  $n_p$  &  $a$  &  $s_{th}$ &  $W$  & $\rho$
\\[2mm]
\hline  \\[-2mm]
 Square  & This work &  12  &  1.05(1)  &  0.43153(3)  &  1.53961(5)
&  0.9994 \\[2mm]
 Square  &  Lieb\cite{li67} &  &  & 0.431523  & 1.539601  &   \\[2mm]
\hline  \\
 Ih   & This work &   9  &  1.84(2)  &  0.41069(8)  &  1.50786(12) &
0.9994 \\[2mm]
 Ih   &  Nagle\cite{na66}  &  &  &  0.41002(10) & 1.50685(15) &
\\[2mm]
 Ih   &  Berg {\em et al.}\cite{be07} &  &  & 0.4104(2)  & 1.5074(3)
&  \\[2mm]
\hline  \\
  VI     & This work &   9  &  3.44(6)  &  0.42138(11) &  1.52406(16)
&  0.9990 \\[2mm]
\hline  \\
 Generic  & Pauling\cite{pa35}  &  &  &  0.405465  &  1.5  &
\\[2mm]
\hline
  \hspace{1.5cm} &   \hspace{1.5cm} &  \hspace{1.0cm} &
\hspace{2.0cm} &
   \hspace{2.5cm} & \hspace{2.5cm} & \hspace{2.0cm} \\
\end{tabular}
\end{table*}

An argument why the entropy per site should decrease for increasing
size is the following. Let us call $\Omega_N$ the number of
configurations compatible with the ice rules for size $N$. For two
independent cells of size $N$ the number of possible configurations
would be $\Omega_N^2$. Then, putting both cells together to form a
larger cell of size $2 N$, we have $\Omega_{2N} < \Omega_N^2$, because
one has to discard configurations that do not ``match'' correctly 
in the border between both $N$-size cells.
The decrease of $s_N$ for increasing $N$ can thus be viewed as an effect of
the boundary conditions.
Assuming that $s_N$ behaves regularly as a function of $1/N$ in the
thermodynamic limit ($1/N \to 0$), one expects for large $N$ a
dependence of the form $s_N = s_{th} + a/N + b/N^2 + ...$. This is in
fact what we find from our simulations, with the parameter $b$ so small
that the linear dependence of $s_N$ on $1/N$ is clearly consistent with 
the results at least for $N > 150$.

\begin{figure}
\vspace{-1.1cm}
\hspace{-0.5cm}
\includegraphics[width= 9cm]{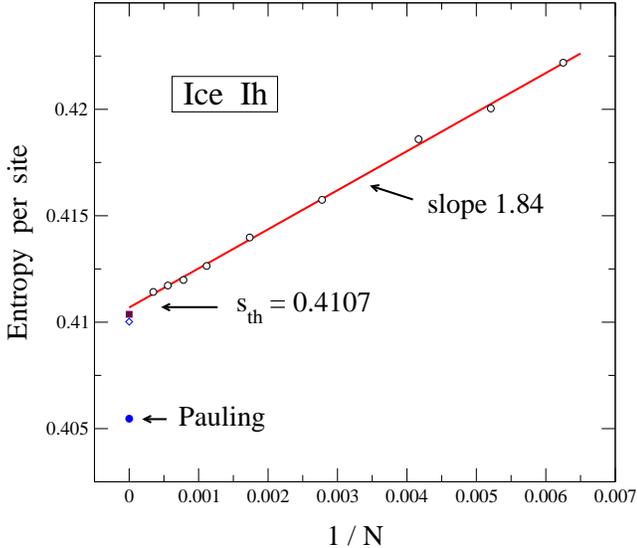}
\vspace{-0.7cm}
\caption{
Entropy per site as a function of the inverse network size ($1/N$)
for
hexagonal ice Ih.
Open circles show results of our thermodynamic integration in the
limit $T \to 0$.  Error bars of the data points are on the order of
the symbol size.
Other symbols represent earlier results for the configurational
entropy of ice Ih:
solid circle, Pauling\cite{pa35}; open diamond, Nagle\cite{na66};
solid square, Berg {\em et al.}\cite{be07}
}
\label{f3}
\end{figure}

For the three-dimensional structure of ice Ih we also find a linear
dependence of the configurational entropy on the inverse network size.
This is shown in Fig.~3, where one observes that the slope (parameter $a$
in Eq.~(\ref{sns})) is larger than in the case of the two-dimensional
lattice. 
The entropy in the thermodynamic limit is lower for ice Ih, as expected 
from earlier results for this ice structure.\cite{na66,be07}
We find $s_{th}$ = 0.41069(8).
As already observed in the analytical result by Nagle\cite{na66} and
in the multicanonical simulations by Berg {\em et al.},\cite{be07}
the configurational entropy is higher than the earlier estimate by
Pauling.
Moreover, our result is slightly higher than that of Nagle\cite{na66},
who found $s_{th}$ = 0.41002.
Our result is a little higher than that found by 
Berg {\em et al.},\cite{be07}, although we consider that both data are
compatible one with the other, given the error bars (see Table~I).

\begin{figure}
\vspace{-1.1cm}
\hspace{-0.5cm}
\includegraphics[width= 9cm]{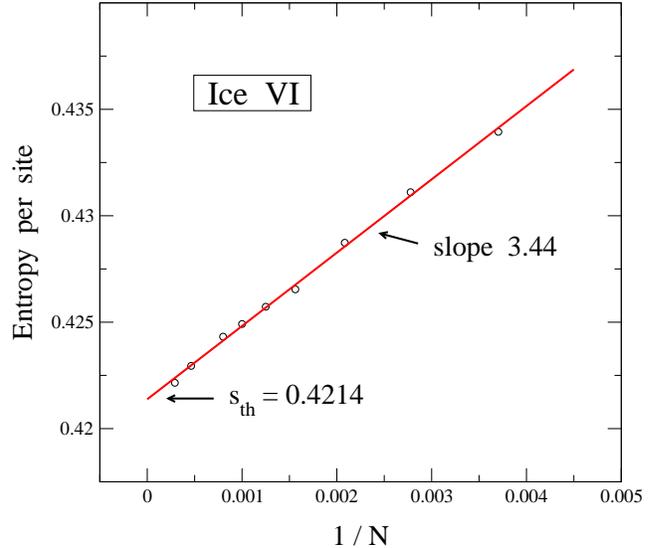}
\vspace{-0.7cm}
\caption{
Entropy per site as a function of the inverse network size ($1/N$)
for ice VI.
Open circles display results of our thermodynamic-integration
procedure in the low-temperature limit.
Error bars are in the order of the symbol size.
}
\label{f4}
\end{figure}

Apart from ice Ih and cubic ice Ic, we are not aware of any 
direct calculation of the configurational entropy of H-disordered ice
for other ice structures. To assess the influence of the ice
network on the entropy, we have also considered the case of ice VI,
where H atoms are known to be disordered.\cite{ku84,pe99,ba12}
In Fig.~4 we display our results for the configurational entropy
of ice VI, as a function of the inverse network size.
We find in this case a slope $a$ = 3.44, much larger than for
the square lattice and ice Ih.
The extrapolation of $s_N$ to infinite network size gives
$s_{th}$ = 0.42138(11), clearly higher than the value corresponding
to ice Ih (a 2.6\% larger). This result is remarkable, since up to
now the only comparison in this respect concerned ice Ih with the
Pauling estimate, being the entropy an 1.3\% larger in the former 
case than in the latter. Now we see that for ice VI the
configurational entropy for this H-disordered structure is 3.9\% larger
than the Pauling result. 

We note the tendency of the entropy $s_{th}$ to increase due to the
presence of loops in the ice structure. 
In fact, the Pauling approximation neglects the presence of loops, as
happens in the so-called Bethe lattice 
(also known as Cayley tree).\cite{zi79,be89}
This gives a value $s_{th} = 0.40547$. 
For ice Ih, which contains six-membered rings of water molecules,
the entropy is higher by an 1.3\%, and it is still
higher for the square lattice with four-membered rings
(a 6.4\% respect the Pauling approach).
With this trend in mind, we could expect for ice VI a value of the 
entropy intermediate between those of ice Ih and the square lattice, 
as it contains four- and six-membered rings 
(apart from other larger loops).\cite{pe99,sa11,si12}
Similarly, for ice networks with larger ring sizes, one can expect
a smaller configurational entropy for disordered hydrogen distributions,
and thus closer to the Pauling result.

Some further comments on Table~I are in order.
In most papers dealing with the configurational entropy of ice,
it is the parameter $W$ which is given, instead of the entropy
$s_{th}$ itself. 
As we find directly $s_{th}$ from our thermodynamic integration,
we have calculated $W$ in the different cases as $W = \exp (s_{th})$.
Also the error bars $\Delta W$ and $\Delta s_{th}$ in the values of 
$s_{th}$ and $W$, respectively, are related by the expression 
$\Delta W = W \, \Delta s_{th}$, as can be derived by differentiating 
the exponential function.
Our error bars for $s_{th}$ represent one standard deviation, as 
given by the least-square procedure employed to fit the data.
In the result by Nagle,\cite{na66} the error bar was estimated by 
this author from an extrapolation of the series terms calculated in 
his analytical procedure.
Note also that, although the result obtained by Pauling was intended to
reproduce the residual entropy of ice Ih, it does not take into account 
the actual ice network, but only the fourfold coordination of the
structure. For this reason we qualify it as ``generic''.
Concerning the results by Berg {\em et al.}\cite{be07}, we note that
these authors have recently\cite{be12} given an entropy value for ice Ih 
slightly smaller than their earlier result, but both of them are compatible 
one with the other, taking into account the statistical error bars. 

To avoid any possible confusion, we emphasize that the simple model
employed in our calculations is not intended to reproduce any physical
characteristic of ice (such as order/disorder transitions) further
than calculating the entropy of an H distribution compatible with the
ice rules.
It implicitly assumes that the distribution of H atoms on the ice
network has no long-range order, and only imposes strict fulfillment 
of the ice rules (short-range order) in the limit $T \to 0$.
The configurational entropy obtained in this way is what has been 
traditionally called residual entropy of ice.\cite{pa35}
With this in mind, we are not allowing for any violation of the third 
law of Thermodynamics. 

In this context, it is generally accepted that the equilibrium ice
phases in the low-temperature limit (at low and high pressures)
display ordered proton structures, as expected from a vanishing of the
entropy.
Order-disorder transitions have been observed between several pairs of
ice phases.\cite{du10,ba12,si12} These transitions are accompanied by
orientational ordering of the water molecules, which implies a
kinetically unfavorable reorganization of the H-bond network. 
In several cases, the transition from a disordered to an ordered phase 
only occurs after doping the sample, which seems to provide a mechanism 
favoring the rearrangement of the H-bond network.\cite{le85,li96,si12}
Thus, at low temperature ice Ih transforms into ice XI, an 
H-ordered phase, but this transition has never been observed in pure
ice, only in doped materials at 72 K.\cite{le85,ho89,li96} 
Something similar happens for high-pressure phases, and ice VI
in particular transforms into proton-ordered ice XV.\cite{sa11,ba12}

Given the entropy difference found here for real structures
such as ice Ih and ice VI, a detailed knowledge of the configurational 
entropy for the different ice phases can be important for precise 
calculations of the phase diagram of water,\cite{sa04,ra12b} 
as noted earlier for ice phases with partial
proton ordering.\cite{ma04b}

\section{Conclusions}

We have presented results for the configurational entropy of
H-disordered ice structures, calculated by means of a thermodynamic 
integration.  A simple model allowed us to derive the
entropy corresponding to each structure.
This procedure has turned out to be very precise, indicating that 
the associated error bars can be made very small without employing 
sophisticated methods, but only using standard statistical mechanics 
and numerical procedures.

For real ice structures, such as ice Ih and ice VI, we find
a difference in configurational entropy of 2.6\%.
For ice VI we obtain an entropy value 3.9\% higher than the Pauling 
estimate, as a consequence of its particular network connectivity. 
This method can be applied to other ice structures with hydrogen
disorder, which will presumably give different values for the 
configurational entropy, due to differences in their network topology.

\begin{acknowledgments}
This work was supported by 
Direcci\'on General de Investigaci\'on Cient\'{\i}fica y T\'ecnica (Spain) 
through Grant FIS2012-31713 
and by Comunidad Aut\'onoma de Madrid 
through Program MODELICO-CM/S2009ESP-1691.
\end{acknowledgments}


\begin{thebibliography}{31}
\expandafter\ifx\csname natexlab\endcsname\relax\def\natexlab#1{#1}\fi
\expandafter\ifx\csname bibnamefont\endcsname\relax
  \def\bibnamefont#1{#1}\fi
\expandafter\ifx\csname bibfnamefont\endcsname\relax
  \def\bibfnamefont#1{#1}\fi
\expandafter\ifx\csname citenamefont\endcsname\relax
  \def\citenamefont#1{#1}\fi
\expandafter\ifx\csname url\endcsname\relax
  \def\url#1{\texttt{#1}}\fi
\expandafter\ifx\csname urlprefix\endcsname\relax\def\urlprefix{URL }\fi
\providecommand{\bibinfo}[2]{#2}
\providecommand{\eprint}[2][]{\url{#2}}

\bibitem[{\citenamefont{Dunaeva et~al.}(2010)\citenamefont{Dunaeva, Antsyshkin,
  and Kuskov}}]{du10}
\bibinfo{author}{\bibfnamefont{A.~N.} \bibnamefont{Dunaeva}},
  \bibinfo{author}{\bibfnamefont{D.~V.} \bibnamefont{Antsyshkin}},
  \bibnamefont{and} \bibinfo{author}{\bibfnamefont{O.~L.}
  \bibnamefont{Kuskov}}, \bibinfo{journal}{Solar System Research}
  \textbf{\bibinfo{volume}{44}}, \bibinfo{pages}{202} (\bibinfo{year}{2010}).

\bibitem[{\citenamefont{Bartels-Rausch
  et~al.}(2012)\citenamefont{Bartels-Rausch, Bergeron, Cartwright, Escribano,
  Finney, Grothe, Gutierrez, Haapala, Kuhs, Pettersson et~al.}}]{ba12}
\bibinfo{author}{\bibfnamefont{T.}~\bibnamefont{Bartels-Rausch}},
  \bibinfo{author}{\bibfnamefont{V.}~\bibnamefont{Bergeron}},
  \bibinfo{author}{\bibfnamefont{J.~H.~E.} \bibnamefont{Cartwright}},
  \bibinfo{author}{\bibfnamefont{R.}~\bibnamefont{Escribano}},
  \bibinfo{author}{\bibfnamefont{J.~L.} \bibnamefont{Finney}},
  \bibinfo{author}{\bibfnamefont{H.}~\bibnamefont{Grothe}},
  \bibinfo{author}{\bibfnamefont{P.~J.} \bibnamefont{Gutierrez}},
  \bibinfo{author}{\bibfnamefont{J.}~\bibnamefont{Haapala}},
  \bibinfo{author}{\bibfnamefont{W.~F.} \bibnamefont{Kuhs}},
  \bibinfo{author}{\bibfnamefont{J.~B.~C.} \bibnamefont{Pettersson}},
  \bibnamefont{et~al.}, \bibinfo{journal}{Rev. Mod. Phys.}
  \textbf{\bibinfo{volume}{84}}, \bibinfo{pages}{885} (\bibinfo{year}{2012}).

\bibitem[{\citenamefont{Eisenberg and Kauzmann}(1969)}]{ei69}
\bibinfo{author}{\bibfnamefont{D.}~\bibnamefont{Eisenberg}} \bibnamefont{and}
  \bibinfo{author}{\bibfnamefont{W.}~\bibnamefont{Kauzmann}},
  \emph{\bibinfo{title}{The Structure and Properties of Water}}
  (\bibinfo{publisher}{Oxford University Press}, \bibinfo{address}{New York},
  \bibinfo{year}{1969}).

\bibitem[{\citenamefont{Petrenko and Whitworth}(1999)}]{pe99}
\bibinfo{author}{\bibfnamefont{V.~F.} \bibnamefont{Petrenko}} \bibnamefont{and}
  \bibinfo{author}{\bibfnamefont{R.~W.} \bibnamefont{Whitworth}},
  \emph{\bibinfo{title}{Physics of Ice}} (\bibinfo{publisher}{Oxford University
  Press}, \bibinfo{address}{New York}, \bibinfo{year}{1999}).

\bibitem[{\citenamefont{Robinson et~al.}(1996)\citenamefont{Robinson, Zhu,
  Singh, and Evans}}]{ro96}
\bibinfo{author}{\bibfnamefont{G.~W.} \bibnamefont{Robinson}},
  \bibinfo{author}{\bibfnamefont{S.~B.} \bibnamefont{Zhu}},
  \bibinfo{author}{\bibfnamefont{S.}~\bibnamefont{Singh}}, \bibnamefont{and}
  \bibinfo{author}{\bibfnamefont{M.~W.} \bibnamefont{Evans}},
  \emph{\bibinfo{title}{Water in Biology, Chemistry and Physics}}
  (\bibinfo{publisher}{World Scientific}, \bibinfo{address}{Singapore},
  \bibinfo{year}{1996}).

\bibitem[{\citenamefont{Bernal and Fowler}(1933)}]{be33}
\bibinfo{author}{\bibfnamefont{J.~D.} \bibnamefont{Bernal}} \bibnamefont{and}
  \bibinfo{author}{\bibfnamefont{R.~H.} \bibnamefont{Fowler}},
  \bibinfo{journal}{J. Chem. Phys.} \textbf{\bibinfo{volume}{1}},
  \bibinfo{pages}{515} (\bibinfo{year}{1933}).

\bibitem[{\citenamefont{Pauling}(1935)}]{pa35}
\bibinfo{author}{\bibfnamefont{L.}~\bibnamefont{Pauling}}, \bibinfo{journal}{J.
  Am. Chem. Soc.} \textbf{\bibinfo{volume}{57}}, \bibinfo{pages}{2680}
  (\bibinfo{year}{1935}).

\bibitem[{\citenamefont{Giauque and Stout}(1936)}]{gi36}
\bibinfo{author}{\bibfnamefont{W.~F.} \bibnamefont{Giauque}} \bibnamefont{and}
  \bibinfo{author}{\bibfnamefont{J.~W.} \bibnamefont{Stout}},
  \bibinfo{journal}{J. Amer. Chem. Soc.} \textbf{\bibinfo{volume}{58}},
  \bibinfo{pages}{1144} (\bibinfo{year}{1936}).

\bibitem[{\citenamefont{Haida et~al.}(1974)\citenamefont{Haida, Matsuo, Suga,
  and Seki}}]{ha74}
\bibinfo{author}{\bibfnamefont{O.}~\bibnamefont{Haida}},
  \bibinfo{author}{\bibfnamefont{T.}~\bibnamefont{Matsuo}},
  \bibinfo{author}{\bibfnamefont{H.}~\bibnamefont{Suga}}, \bibnamefont{and}
  \bibinfo{author}{\bibfnamefont{S.}~\bibnamefont{Seki}}, \bibinfo{journal}{J.
  Chem. Thermodyn.} \textbf{\bibinfo{volume}{6}}, \bibinfo{pages}{815}
  (\bibinfo{year}{1974}).

\bibitem[{\citenamefont{Nagle}(1966)}]{na66}
\bibinfo{author}{\bibfnamefont{J.~F.} \bibnamefont{Nagle}},
  \bibinfo{journal}{J. Math. Phys.} \textbf{\bibinfo{volume}{7}},
  \bibinfo{pages}{1484} (\bibinfo{year}{1966}).

\bibitem[{\citenamefont{Berg et~al.}(2007)\citenamefont{Berg, Muguruma, and
  Okamoto}}]{be07}
\bibinfo{author}{\bibfnamefont{B.~A.} \bibnamefont{Berg}},
  \bibinfo{author}{\bibfnamefont{C.}~\bibnamefont{Muguruma}}, \bibnamefont{and}
  \bibinfo{author}{\bibfnamefont{Y.}~\bibnamefont{Okamoto}},
  \bibinfo{journal}{Phys. Rev. B} \textbf{\bibinfo{volume}{75}},
  \bibinfo{pages}{092202} (\bibinfo{year}{2007}).

\bibitem[{\citenamefont{Howe and Whitworth}(1987)}]{ho87}
\bibinfo{author}{\bibfnamefont{R.}~\bibnamefont{Howe}} \bibnamefont{and}
  \bibinfo{author}{\bibfnamefont{R.~W.} \bibnamefont{Whitworth}},
  \bibinfo{journal}{J. Chem. Phys.} \textbf{\bibinfo{volume}{86}},
  \bibinfo{pages}{6443} (\bibinfo{year}{1987}).

\bibitem[{\citenamefont{MacDowell et~al.}(2004)\citenamefont{MacDowell, Sanz,
  Vega, and Abascal}}]{ma04b}
\bibinfo{author}{\bibfnamefont{L.~G.} \bibnamefont{MacDowell}},
  \bibinfo{author}{\bibfnamefont{E.}~\bibnamefont{Sanz}},
  \bibinfo{author}{\bibfnamefont{C.}~\bibnamefont{Vega}}, \bibnamefont{and}
  \bibinfo{author}{\bibfnamefont{J.~L.~F.} \bibnamefont{Abascal}},
  \bibinfo{journal}{J. Chem. Phys.} \textbf{\bibinfo{volume}{121}},
  \bibinfo{pages}{10145} (\bibinfo{year}{2004}).

\bibitem[{\citenamefont{Berg and Yang}(2007)}]{be07b}
\bibinfo{author}{\bibfnamefont{B.~A.} \bibnamefont{Berg}} \bibnamefont{and}
  \bibinfo{author}{\bibfnamefont{W.}~\bibnamefont{Yang}}, \bibinfo{journal}{J.
  Chem. Phys.} \textbf{\bibinfo{volume}{127}}, \bibinfo{pages}{224502}
  (\bibinfo{year}{2007}).

\bibitem[{\citenamefont{Isakov et~al.}(2004)\citenamefont{Isakov, Raman,
  Moessner, and Sondhi}}]{is04}
\bibinfo{author}{\bibfnamefont{S.~V.} \bibnamefont{Isakov}},
  \bibinfo{author}{\bibfnamefont{K.~S.} \bibnamefont{Raman}},
  \bibinfo{author}{\bibfnamefont{R.}~\bibnamefont{Moessner}}, \bibnamefont{and}
  \bibinfo{author}{\bibfnamefont{S.~L.} \bibnamefont{Sondhi}},
  \bibinfo{journal}{Phys. Rev. B} \textbf{\bibinfo{volume}{70}},
  \bibinfo{pages}{104418} (\bibinfo{year}{2004}).

\bibitem[{\citenamefont{Ziman}(1979)}]{zi79}
\bibinfo{author}{\bibfnamefont{J.~M.} \bibnamefont{Ziman}},
  \emph{\bibinfo{title}{Models of disorder}} (\bibinfo{publisher}{Cambridge
  University}, \bibinfo{address}{Cambridge}, \bibinfo{year}{1979}).

\bibitem[{\citenamefont{Bell and Lavis}(1989)}]{be89}
\bibinfo{author}{\bibfnamefont{G.~M.} \bibnamefont{Bell}} \bibnamefont{and}
  \bibinfo{author}{\bibfnamefont{D.~A.} \bibnamefont{Lavis}},
  \emph{\bibinfo{title}{Statistical Mechanics of Lattice Models. Volume 1:
  Closed Form and Exact Theories of Cooperative Phenomena}}
  (\bibinfo{publisher}{Ellis Horwood Ltd.}, \bibinfo{address}{New York},
  \bibinfo{year}{1989}).

\bibitem[{\citenamefont{Lieb}(1967{\natexlab{a}})}]{li67}
\bibinfo{author}{\bibfnamefont{E.~H.} \bibnamefont{Lieb}},
  \bibinfo{journal}{Phys. Rev. Lett.} \textbf{\bibinfo{volume}{18}},
  \bibinfo{pages}{692} (\bibinfo{year}{1967}{\natexlab{a}}).

\bibitem[{\citenamefont{Lieb}(1967{\natexlab{b}})}]{li67b}
\bibinfo{author}{\bibfnamefont{E.~H.} \bibnamefont{Lieb}},
  \bibinfo{journal}{Phys. Rev.} \textbf{\bibinfo{volume}{162}},
  \bibinfo{pages}{162} (\bibinfo{year}{1967}{\natexlab{b}}).

\bibitem[{\citenamefont{Chandler}(1987)}]{ch87}
\bibinfo{author}{\bibfnamefont{D.}~\bibnamefont{Chandler}},
  \emph{\bibinfo{title}{Introduction to modern statistical mechanics}}
  (\bibinfo{publisher}{Oxford University Press}, \bibinfo{address}{Oxford},
  \bibinfo{year}{1987}).

\bibitem[{\citenamefont{Binder and Heermann}(1997)}]{bi97}
\bibinfo{author}{\bibfnamefont{K.}~\bibnamefont{Binder}} \bibnamefont{and}
  \bibinfo{author}{\bibfnamefont{D.~W.} \bibnamefont{Heermann}},
  \emph{\bibinfo{title}{Monte Carlo Simulation in Statistical Physics}}
  (\bibinfo{publisher}{Springer}, \bibinfo{address}{Berlin},
  \bibinfo{year}{1997}), \bibinfo{edition}{3rd} ed.

\bibitem[{\citenamefont{Herrero and Ram\'{\i}rez}(1992)}]{he92}
\bibinfo{author}{\bibfnamefont{C.~P.} \bibnamefont{Herrero}} \bibnamefont{and}
  \bibinfo{author}{\bibfnamefont{R.}~\bibnamefont{Ram\'{\i}rez}},
  \bibinfo{journal}{Chem. Phys. Lett.} \textbf{\bibinfo{volume}{194}},
  \bibinfo{pages}{79} (\bibinfo{year}{1992}).

\bibitem[{\citenamefont{Kuhs et~al.}(1984)\citenamefont{Kuhs, Finney, Vettier,
  and Bliss}}]{ku84}
\bibinfo{author}{\bibfnamefont{W.~F.} \bibnamefont{Kuhs}},
  \bibinfo{author}{\bibfnamefont{J.~L.} \bibnamefont{Finney}},
  \bibinfo{author}{\bibfnamefont{C.}~\bibnamefont{Vettier}}, \bibnamefont{and}
  \bibinfo{author}{\bibfnamefont{D.~V.} \bibnamefont{Bliss}},
  \bibinfo{journal}{J. Chem. Phys.} \textbf{\bibinfo{volume}{81}},
  \bibinfo{pages}{3612} (\bibinfo{year}{1984}).

\bibitem[{\citenamefont{Salzmann et~al.}(2011)\citenamefont{Salzmann, Radaelli,
  B, and L}}]{sa11}
\bibinfo{author}{\bibfnamefont{C.~G.} \bibnamefont{Salzmann}},
  \bibinfo{author}{\bibfnamefont{P.~G.} \bibnamefont{Radaelli}},
  \bibinfo{author}{\bibfnamefont{S.}~\bibnamefont{B}}, \bibnamefont{and}
  \bibinfo{author}{\bibfnamefont{F.~J.} \bibnamefont{L}},
  \bibinfo{journal}{Phys. Chem. Chem. Phys.} \textbf{\bibinfo{volume}{13}},
  \bibinfo{pages}{18468} (\bibinfo{year}{2011}).

\bibitem[{\citenamefont{Singer and Knight}(2012)}]{si12}
\bibinfo{author}{\bibfnamefont{S.~J.} \bibnamefont{Singer}} \bibnamefont{and}
  \bibinfo{author}{\bibfnamefont{C.}~\bibnamefont{Knight}},
  \bibinfo{journal}{Adv. Chem. Phys.} \textbf{\bibinfo{volume}{147}},
  \bibinfo{pages}{1} (\bibinfo{year}{2012}).

\bibitem[{\citenamefont{Berg et~al.}(2012)\citenamefont{Berg, Muguruma, and
  Okamoto}}]{be12}
\bibinfo{author}{\bibfnamefont{B.~A.} \bibnamefont{Berg}},
  \bibinfo{author}{\bibfnamefont{C.}~\bibnamefont{Muguruma}}, \bibnamefont{and}
  \bibinfo{author}{\bibfnamefont{Y.}~\bibnamefont{Okamoto}},
  \bibinfo{journal}{Mol. Sim.} \textbf{\bibinfo{volume}{38}},
  \bibinfo{pages}{856} (\bibinfo{year}{2012}).

\bibitem[{\citenamefont{Leadbetter et~al.}(1985)\citenamefont{Leadbetter, Ward,
  Clark, Tucker, Matsuo, , and Suga}}]{le85}
\bibinfo{author}{\bibfnamefont{A.~J.} \bibnamefont{Leadbetter}},
  \bibinfo{author}{\bibfnamefont{R.~C.} \bibnamefont{Ward}},
  \bibinfo{author}{\bibfnamefont{J.~W.} \bibnamefont{Clark}},
  \bibinfo{author}{\bibfnamefont{P.~A.} \bibnamefont{Tucker}},
  \bibinfo{author}{\bibfnamefont{T.}~\bibnamefont{Matsuo}}, , \bibnamefont{and}
  \bibinfo{author}{\bibfnamefont{H.}~\bibnamefont{Suga}}, \bibinfo{journal}{J.
  Chem. Phys.} \textbf{\bibinfo{volume}{82}}, \bibinfo{pages}{424}
  (\bibinfo{year}{1985}).

\bibitem[{\citenamefont{Line and Whitworth}(1996)}]{li96}
\bibinfo{author}{\bibfnamefont{C.~M.~B.} \bibnamefont{Line}} \bibnamefont{and}
  \bibinfo{author}{\bibfnamefont{R.~W.} \bibnamefont{Whitworth}},
  \bibinfo{journal}{J. Chem. Phys.} \textbf{\bibinfo{volume}{104}},
  \bibinfo{pages}{10008} (\bibinfo{year}{1996}).

\bibitem[{\citenamefont{Howe and Whitworth}(1989)}]{ho89}
\bibinfo{author}{\bibfnamefont{R.}~\bibnamefont{Howe}} \bibnamefont{and}
  \bibinfo{author}{\bibfnamefont{R.~W.} \bibnamefont{Whitworth}},
  \bibinfo{journal}{J. Chem. Phys.} \textbf{\bibinfo{volume}{90}},
  \bibinfo{pages}{4450} (\bibinfo{year}{1989}).

\bibitem[{\citenamefont{Sanz et~al.}(2004)\citenamefont{Sanz, Vega, Abascal,
  and MacDowell}}]{sa04}
\bibinfo{author}{\bibfnamefont{E.}~\bibnamefont{Sanz}},
  \bibinfo{author}{\bibfnamefont{C.}~\bibnamefont{Vega}},
  \bibinfo{author}{\bibfnamefont{J.~L.~F.} \bibnamefont{Abascal}},
  \bibnamefont{and} \bibinfo{author}{\bibfnamefont{L.~G.}
  \bibnamefont{MacDowell}}, \bibinfo{journal}{Phys. Rev. Lett.}
  \textbf{\bibinfo{volume}{92}}, \bibinfo{pages}{255701}
  (\bibinfo{year}{2004}).

\bibitem[{\citenamefont{Ram\'{\i}rez et~al.}(2012)\citenamefont{Ram\'{\i}rez,
  Neuerburg, and Herrero}}]{ra12b}
\bibinfo{author}{\bibfnamefont{R.}~\bibnamefont{Ram\'{\i}rez}},
  \bibinfo{author}{\bibfnamefont{N.}~\bibnamefont{Neuerburg}},
  \bibnamefont{and} \bibinfo{author}{\bibfnamefont{C.~P.}
  \bibnamefont{Herrero}}, \bibinfo{journal}{J. Chem. Phys.}
  \textbf{\bibinfo{volume}{137}}, \bibinfo{pages}{134503}
  (\bibinfo{year}{2012}).

\end{thebibliography}
\end{document}